# Western ideological homogeneity in entrepreneurial finance research: Evidence from highly cited publications

Minh-Hoang Nguyen [1,2,*]

Huyen Thanh T. Nguyen [3,4]

Thanh-Hang Pham [5,6]

Manh-Toan Ho [1,4]

Quan-Hoang Vuong [1,4]

[1] Centre for Interdisciplinary Social Research, Phenikaa University, Yen Nghia Ward, Ha Dong District, Hanoi 100803, Vietnam

[2] Graduate School of Asia Pacific Studies, Ritsumeikan Asia Pacific University, Beppu, Oita 874-8577, Japan

[3] School of Economics and International Business, Foreign Trade University, 91 Chua Lang, Dong Da, Hanoi 100000

[4] AISDL, Vuong & Associates, Hanoi 100000, Vietnam

[5] School of Business, RMIT Vietnam University, Hanoi, 100000, Viet Nam

[6] Faculty of Management and Tourism, Hanoi University, Km9, Nguyen Trai Road, Thanh Xuan, Hanoi 100803, Vietnam

(*) Corresponding author; email: hoang.nguyenminh@phenikaa-uni.edu.vn

## Abstract

Entrepreneurs play crucial roles in global sustainable development, but limited financial resources constrain their performance and survival rate. Entrepreneurial finance discipline is, therefore, born to explore the connection between finance and entrepreneurship. Despite the global presence of entrepreneurship, the literature of entrepreneurial finance is suspected to be Western ideologically homogenous. Thus, the objective of this study is to examine the existence of Western ideological homogeneity in entrepreneurial finance literature. Employing the mindsponge mechanism and bibliometric analyses (Y-index and social structure), we analyze 412 highly cited publications extracted from Web of Science database and find Western ideological dominance as well as weak tolerance towards heterogeneity in the set of core ideologies of entrepreneurial finance. These results are consistent across author-, institution-, and country-levels, which reveals strong evidence for the existence of Western ideological homogeneity in the field. We recommend editors, reviewers, and authors to have proactive actions to diversify research topics and enhancing knowledge exchange to avoid the shortfalls of ideological homogeneity. Moreover, the synthesis of mindsponge mechanism and bibliometric analyses are suggested as a possible way to evaluate the state of ideological diversity in other scientific disciplines.

## 1. Introduction

Entrepreneurship is not only an important engine that drives the economy but also a contributor to sustainable development. Besides creating jobs and fostering innovation in the economic sector, entrepreneurs also join hands to combat social issues around the globe; most recently is the COVID-19 pandemic, which has infected more than 16 million people and resulted in approximately 650 thousand deaths as of 27 July 2020 (Dayton, 2020). Entrepreneurship also supports poverty reduction in emerging countries without compromising environmental quality (Bruton, Ahlstrom, & Si, 2015; Dean & McMullen, 2007; Dhahri & Omri, 2018; McMullen, 2011; Vuong, Ho, Nguyen, & Nguyen, 2019; Vuong, 2014; Vuong et al., 2020). However, the survival rate and performance of startups, especially those in emerging countries, are often affected by financial constraints. Despite the demand for scientific findings regarding financing methods for entrepreneurs in emerging countries, a majority of researches in entrepreneurial finance is based on Western viewpoints. These viewpoints primarily focus on financing sources that associate with advanced technological development, such as venture capital, private equity, crowdfunding, and so forth (Cumming & Groh, 2018). This situation might be the result of Western ideological homogeneity (Nguyen, Nguyen, Pham, Nguyen, & Vuong, 2020). Nevertheless, the evidence for this argument is insufficient; therefore, the current study aims to examine whether the literature of entrepreneurial finance is ideologically homogenous.

While most of the government initiatives and solutions have been found to create a limited impact on reducing poverty, entrepreneurship and new ventures can more or less serve as an effective solution to poverty around the world (Ahlstrom, 2010; Alvarez, Barney, & Newman, 2015; Bruton et al., 2015). However, compared to entrepreneurs in developed economies, entrepreneurs in emerging economies do not just deal with resource constraints but also other obstacles, such as political instability and underdeveloped rule enforcement mechanisms, which hinder the development of venture capital and crowdfunding mechanisms (Scott, Sinha, Gibb, & Akoorie, 2020). Still, research on financing methods other than venture capital and crowdfunding of entrepreneurs in developing countries is limited. One of the most frequently proposed financing methods for entrepreneurs in poor regions is microlending. Nevertheless, microlending only helps entrepreneurs to make ends meet rather than help them to build new businesses (Bruton et al., 2015; Vermeire & Bruton, 2016).

In dynamic economies like China and South-East Asian countries, entrepreneurial finance research mostly concentrates on Western-based financing sources rather than locally conventional sources. For example, the venture capital industry in China only started to develop after the Chinese government established policies to encourage venture investments in 1998 (Batjargal & Liu, 2004). Since then, research about entrepreneurial finance in China has been mostly about venture capital (Nguyen, Nguyen, et al., 2020) but virtually neglected other China's culture-based and advantage-based financing methods, such as family financing and governmental subsidies, and so forth (Ahlstrom & Ding, 2014).

The tie between entrepreneurship and venture capital, in particular, is indeed a double-edged solution for economic prosperity. Thanks to venture capital, technology startups during the 1990s developed rapidly, which fueled the Internet revolution. From the 1980s to 1999, venture capital fund in the USA rocketed from more than 5 billion dollars to roughly $70 billion. Around 60% of the investment went to information technology industries (e.g., communications and networking, software, and information services), and around 10% went to life sciences (e.g., biotechnology) sector (Gompers & Lerner, 2001). One of the most significant symbols for the success resulting from the relationship between venture capital and entrepreneurship was the "miracle" of Silicon Valley (Ferrary & Granovetter, 2009).

Nonetheless, the motivation of venture capitalists to pursue and maximize profit was not only the accelerator of the internet revolution but also the magnet pulling them to the illusion of a new economic era with "endless" growth. That illusion eventually led to the "dot-com bubble" in 2000. After the crisis, the market value of Internet enterprises plunged from $1 trillion in March 2000 to $572 billion in December. At the same time, almost 800 Internet enterprises faded (Goodnight & Green, 2010). Thus, it is plausible to say that the overreliance on a financing method may not be financially sustainable for entrepreneurs, and the rule of diversification should not be violated, even in the scholarly aspect. In scientific research, Nguyen, Nguyen, et al. (2020) indicate a sign of Western ideological homogeneity in the literature of entrepreneurial finance and call for firm evidence to support the claim.

Thus, we aim to examine whether there is an existence of Western ideological homogeneity in the entrepreneurial discipline. To the best of our knowledge, no prior studies have been conducted to assess the ideological homogeneity/diversity of a scientific discipline, so we attempt to propose a new method using bibliometric analysis and the mindsponge mechanism (Vuong & Napier, 2015) for doing so.

## 2. Theoretical background

### 2.1. Ideology and how to identify it

The definition of ideology is myriad, and there is currently no general agreement on its definition. The origin of 'ideology' started more than 200 years ago when it was first coined by the French philosopher Destutt de Tracy to indicate a new discipline that would study 'ideas': *idéologie* (Van Dijk, 2006). Since then, a significant number of variations on the definition of ideology have been circulating within the social sciences under different contexts and scenarios (Gerring, 1997). For example, socialists describe ideology as "cultural beliefs that justify particular social arrangements, including patterns of inequality" (Macionis & Gerber, 2010). Meanwhile, political scientists define ideology as "a set of ideas, beliefs, values, and opinions, exhibiting a recurring pattern, that competes deliberately as well as unintentionally over providing plans of action for public policymaking in an attempt to justify, explain, contest, or change the social and political arrangements and processes of a political community" (Freeden, 2001).

The existence of various definitions regarding ideology, indeed, makes the determination of an appropriate definition of ideology complicated. However, there exists one commonly accepted core definition: "a set of idea-elements that are bound together, that belong to one another in a non-random fashion," argued by Gerring (1997). Moreover, as academia is formed by myriad scientific societies globally, the selected definition should highly present the socialness. We, thus, refer to the definition posed by Van Dijk (2006) because of its generality and socialness: " ideology is the foundation of the social representations shared by a social group." Eventually, we define an ideology as a set of ideas and beliefs that is shared by a group of researchers.

The ideology can be distinguished by various means, such as socio-cultural, epistemological, ethical, political, geographical, or religious characteristics of a social group. Among those approaches, the classification based on geographical location is one of the most common practices. Based on the geographical location, the institutional, socio-cultural, and economic values of a group of researchers can be differentiated. We acknowledge that the ideology of researchers within a specific geographical area might be different politically, epistemologically, and ethically, but the typical set of beliefs shared by the majority of researchers in the given area makes them differentiated. For instance, a majority of researchers in Western societies (e.g., the USA, UK, Canada, etc.) will have a different ideological viewpoint with their peers in Eastern societies (e.g., China, Japan, Korea, Vietnam, etc.). Therefore, it is plausible to say the affiliations of researchers can more or less represent the ideologies the papers convey.

### 2.2. Ideological homogeneity and how to measure it

The issue of ideological homogeneity has been commonly discussed in social sciences, especially political science, for years. Ideological homogeneity is usually referred to as the state of lacking diverse beliefs and principles in a group of likeminded people (Wojcieszak, 2010). Multiple efforts are made to measure the degree of either ideological homogeneity or heterogeneity. In political science, scientists usually employ demographic and opinion proxies to estimate the level of ideological diversity among constituents or population within a specific geographical, legislative, or social boundary (Bond, 1983; Bullock & Brady, 1983; Sullivan, 1973). To elaborate, the higher variance among opinions of respondents, the higher level of ideological heterogeneity and vice versa (Levendusky & Pope, 2010).

Even though collecting opinion data is expected to provide more advantages than relying on demographic proxies, the method is not applicable in the case of scientific publishing due to infeasibility. The survey distribution to every single researcher in a scientific discipline is not only timely and costly but also ineffective. Many researchers might have changed their affiliation or email address after several years, and not all email addresses are available. Survey design is another significant challenge. We do not focus solely on any type of ideology (e.g., political ideology, cultural ideology, epistemological ideology, etc.) but rather all the ideological facets that can be observed through the geographical location. By this way, the ideological difference across regions can be differentiated through the kind of the civilization of the given region (e.g., Western, Islamic, Buddhist, Hindu, Orthodox, Sinic, African ideologies, etc.) or types of government (e.g., capitalist, communist, Islamic ideologies, etc.). Hence, no current set of questions is adequate to measure such dynamic ideological differences.

In an ideologically homogenous environment, the community/group/population's viewpoint is driven mainly by the dominant ideology, while other ideologies are suppressed. The suppression of other ideologies leads to a low level of tolerance of the community towards various sets of values (Atkeson & Taylor, 2019; Rom, 2019). We, therefore, can measure the level of ideological homogeneity by acknowledging these two primary characteristics (dominance and tolerance) and apply them in the context of scientific publishing. In scientific publishing, counting the number of publications can help measure the prevalence of an ideology in a discipline. However, it is not enough to assess the ideological dominance, because larger quantity does not necessarily represent more considerable influence. For instance, China ranks 4$^{th}$ in scientific production but has a relatively low scientific impact in the entrepreneurial finance discipline (Nguyen, Nguyen, et al., 2020). Therefore, identifying the boundary between highly influential publications and popular publications is necessary to assess whether a scientific discipline is ideologically homogenous or not.

We employ the "mindsponge" mechanism proposed by Vuong and Napier (2015) for better differentiation between highly influential publications and well-known publications as well as assessing the ideological homogeneity. We assume that every scientific discipline has a "nucleus" or a set of ideologies or core values that editors/reviewers/authors use to judge the usefulness of the information or expand the literature upon (see Figure 1). This "nucleus" is elusive, but it can be evaluated by analyzing highly cited publications in the field. By nature, highly cited publications are works that pose significant impact and influence over the thinking of other researchers in the respective field (Hui-Zhen & Ho, 2015), which is similar to the functions of the "mindset" at the individual-level (Vuong, 2016; Vuong & Napier, 2015). When a publication is highly cited, the contained values or ideologies of the given publication are perceived as crucial for the discipline by a large number of researchers who play as trust evaluators. Eventually, the citation system can be considered as the filtering mechanism of a scientific discipline to integrate, synthesize, and incorporate ideologies that are aligned with the "nucleus." The buffer zone surrounding the "nucleus" is constructed by the scholarly works published by qualified journals. In contrast, the utmost marginal zone contains the cultural and ideological values of a particular setting to which the scientific discipline contributes (here we set as global context).

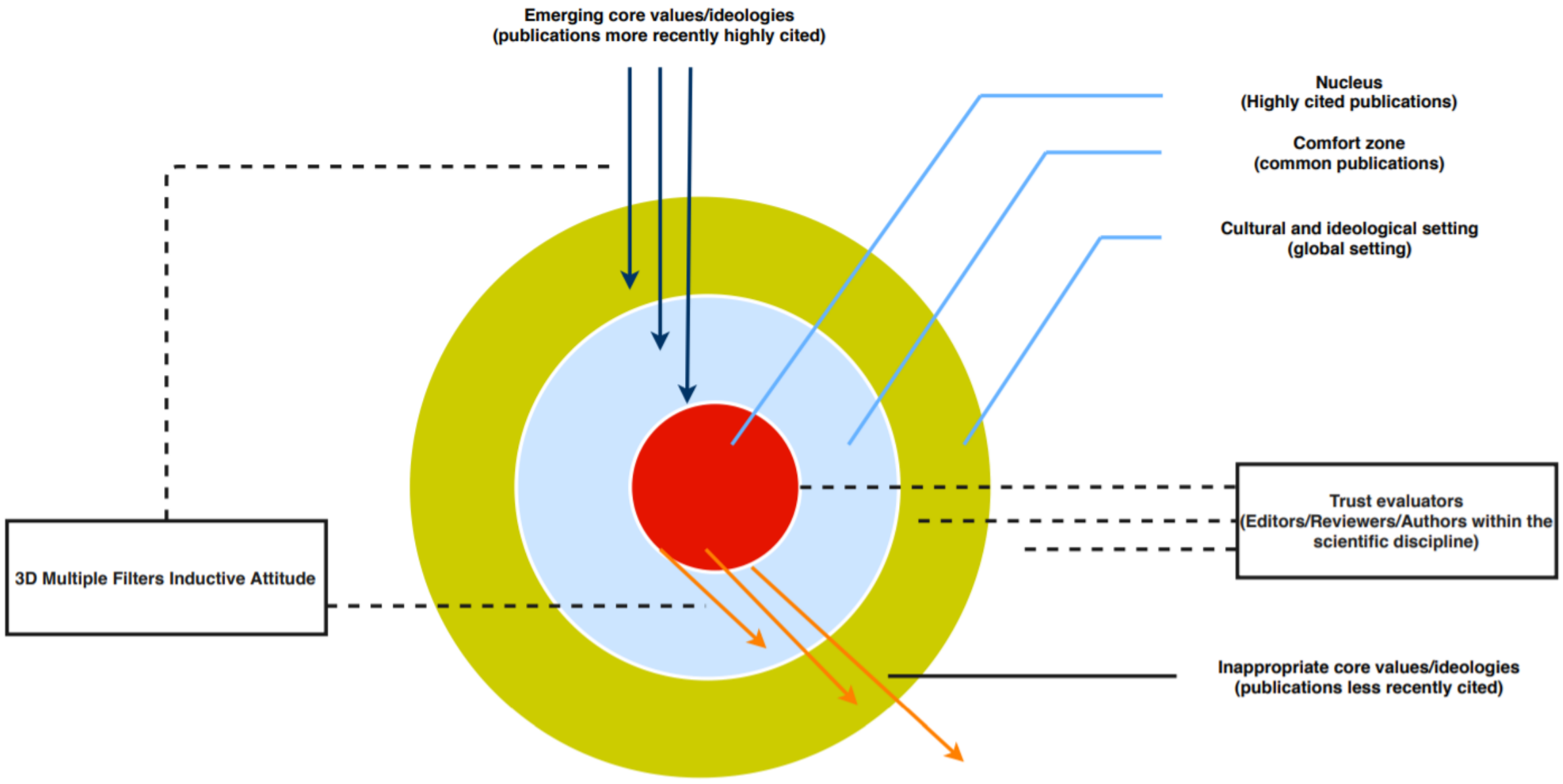

**Figure 1:** The mindsponge mechanism in a scientific discipline, adapted from Vuong & Napier (2015).

Based on the mindsponge mechanism, the *dominance* of an ideology can be measured by counting the number of publications in the "nucleus" or highly cited publications. The higher prevalence of highly cited publications with a similar ideology, the more dominant the given ideology is within the scientific discipline. The ideological dominance alone is not enough to represent the homogeneity, as it lacks an indication of "a group of like-minded people." In an ideologically homogenous group, people tend to "suppress alternative viewpoints, and encourage the self-censorship of deviant ideas" to avoid conflict (Atkeson & Taylor, 2019). Therefore, a supplementary evaluation indicator of ideological homogeneity is the discipline's *tolerance* towards heterogeneity. We define tolerance as the degree that the scientific discipline accepts the coexistence of the dominant ideology with other different ideologies in the "nucleus." In sum, if the proportion of non-dominant ideologies within both the collection of highly cited publications and the boundary of collaboration networks is low or absent, the field can be considered as ideological homogenous, and vice versa.

## 3. Methods and materials

To justify the Western ideological homogeneity in the entrepreneurial finance literature, we focus on evaluating two matters: i) the *dominance* of Western ideology among highly cited publications, and ii) the discipline's *tolerance* of other ideologies other than Western ideology among highly cited publications. The Y-index is employed to assess *dominance*, while co-authorship analysis is employed to visualize the social structure for evaluating *tolerance*. Both techniques are conducted across three levels of a publication (author, institution, country levels) for acquiring different views from the big picture to a finer scale.

### 3.1. Bibliometrics analysis

#### 3.1.1. Y-index

Usually, the productivity (or scientific relevance) of an author is justified based on his/her number of publications using full counting or fractional counting. However, both metrics are not suitable in the current analysis. Full counting gives each of the N authors full credit of a publication, which is convenient but neglects the real contribution of the author and inflates the productivity of authors with high collaboration tendency (Huang, Lin, & Chen, 2011). Fractional counting gives a partial credit of 1/N to each of N authors in a publication, which is a seemingly fairer approach to evaluate the contribution of an author than the full counting. Nevertheless, both counting methods fail to address the leadership and conceptualization roles, which are essential to assess the ideological *dominance* of an author/institution/country over the article. Thus, the Y-index is selected for its advantages, such as revealing fundamental contributions (or leadership), ignoring unethical practices (e.g., gift authorship), and providing features of contribution (Fu & Ho, 2014).

The Y-index is proposed as a new method to evaluate the performance and characterize the contribution of an author, an institution, or a country. The index is estimated by using the number of first-authored (FP) and corresponding (RP) publications; first author and corresponding author are two most prominent authorship positions in the paper (Mattsson, Sundberg, & Laget, 2011; Riesenberg & Lundberg, 1990). The index has been widely employed in many studies of highly cited papers in multiple fields, such as biomass research, dental research, information, and library science research (Chen & Ho, 2015; Ivanović & Ho, 2016;

Yeung & Ho, 2019). The Y-index is defined through two parameters *j* and *h*, which are calculated by the following formulas, respectively:

$$j = FP + RP$$

$$h = \tan^{-1}\left(\frac{RP}{FP}\right)$$

After the $j$ and $h$ values are obtained, the Y-index can be demonstrated on a two-dimensional polar coordinate with $j \cos h$ being the x-axis and $j \sin h$ being the y-axis. An author with higher *j* will hold a more significant role in the field and will be positioned further away from the origin of the polar coordinate (0, 0). When the author has equal numbers of corresponding publications and first-authored publications, $h = 0.7854$. $h < 0.7854$ indicates the author to obtain more first-author publications, while $h > 0.7854$ indicates the author to obtain more corresponding publications. Notably, $j =$ number of first-author publications when $h = 0$, and $j =$ number of corresponding publications when $h = \frac{\pi}{2}$. The calculation can be similarly applied to institution- and country-levels.

#### 3.1.2. Co-authorship analysis

Co-authorship analysis is a common practice to examine the collaborative activities in a scientific discipline. The analysis documents the interactions among authors to create a co-authorship network or social structure that displays the collaboration patterns of not only authors but also their institutions and countries (Reyes-Gonzalez, Gonzalez-Brambila, & Veloso, 2016). The emphasis of co-authorship analysis is not on the attributes of the authors/institutions/countries but the connections among them in the network system (e Fonseca, Sampaio, de Araújo Fonseca, & Zicker, 2016). This attribute of co-authorship analysis is widely employed to identify key leading and weakly engaged actors in the network as well as the collaboration tendencies of those actors. Visually, the network is constructed from a mixture of nodes and edges. Each node in a network represents an author/institution/country, while an edge established between two nodes represents the connection between two given nodes. The size of a node is proportionate to the total frequency of collaborations of the given nodes with others. In contrast, the size of an edge corresponds to the number of collaborations between two nodes connected by the given edge.

Collaboration has long been considered as a means to exchange knowledge, enhance specialization, and integrate complex information, but it requires a consensus among collaborators to achieve the expected outcomes. Therefore, to gain effective collaboration, like-minded people tend to work together; otherwise, there has to be high tolerance of heterogeneity among group members. Based on this attribute of collaboration, we determine to employ co-authorship analysis for evaluating the *tolerance* of heterogeneity within the "nucleus" of entrepreneurial finance.

### 3.2. Materials

We select the Web of Science (WoS) database as the source of data for this analysis. Governmental agencies and international organizations have used the database, which encompasses a wide range of qualified publications from 1900 to the present to evaluate scientific performance and the impact of scientists, institutions, and countries (Nguyen, Ho et al., 2020).

Entrepreneurial finance is an overlap between entrepreneurship and finance disciplines. Cumming and Johan (2017) assert that entrepreneurial finance literature is so interdisciplinary that it also covers knowledge in disciplines other than entrepreneurship and finance, such as public policy, psychology, sociology, and geography, etc. Therefore, we define entrepreneurial finance as studies that cover both the attributes of entrepreneurship and finance. As such, based on prior pieces of literature in entrepreneurship (Aparicio, Iturralde, & Maseda, 2019; Vallaster, Kraus, Lindahl, & Nielsen, 2019) and finance (Cumming & Groh, 2018; Padilla-Ospina, Medina-Vásquez, & Rivera-Godoy, 2018; Xu et al., 2018; D. Zhang, Zhang, & Managi, 2019), we select two sets of search keywords respectively, and then take their intersection using the 'AND' Boolean.

- ("entrepreneur*" OR "startup*" OR "start-up*" OR "new enterprise*" OR "new firm*")
- ("financ*" OR "debt*" OR "venture capital*" OR "trade credit*" OR "crowdfund*" OR "angel invest*" OR "private equit*" OR "IPO*")

The search was conducted on the 2 March 2020 through the field tag "Topic" without any restriction on publication types or publication period. The only inclusion criterion was that the extracted publications need to be written in English. In total, 10,0529 records were retrieved.

To identify highly cited publications, there are currently two predominant methods. One way is to set a specific citation rate or threshold, whereas another way is to select a specific number of most cited publications (e.g., top 1% publications for the number of citations) (X. Zhang, Estoque, Xie, Murayama, & Ranagalage, 2019). In the current study, we employ the former approach to determine highly cited publications; in detail, we set a citation threshold of more than 100 citations. This threshold is also applied in many other studies (Barbosa & Schneck, 2015; Fu & Ho, 2016; X. Zhang et al., 2019). Moreover, any publications that can receive more than 100 citations are proven to be crucial components of the discipline, so they are all qualified for analysis regardless of publication type. As a result, the publications are not qualified for analysis according to two following exclusion criteria: 1) the publication obtains less than 100 citations, and 2) the publication's authors are anonymous.

### 3.3. Procedure

The analysis in this study is separated into several steps. First, the data is extracted from the WoS database using the aforementioned search keywords and saved as 'csv.' and 'txt.' files. Second, we apply the exclusion criteria to exclude all unqualified publications. Third, we manually disambiguate the authors' names and computationally disambiguate authors' affiliations. For example, 'Cumming D', 'Cumming DJ', and 'Cumming Douglas' are one single author, but the software will interpret them as different authors if the manual disambiguation process is not conducted. Forth, the first authors' and corresponding authors' names and affiliations are generated in Excel to calculate the Y-index. Lastly, the co-authorship analysis is implemented using the **bibliometrix** R package (Aria & Cuccurullo, 2017). Limitations of the current study are also discussed in the Discussion for transparency (Vuong, 2020).

## 4. Result

### 4.1. Overview

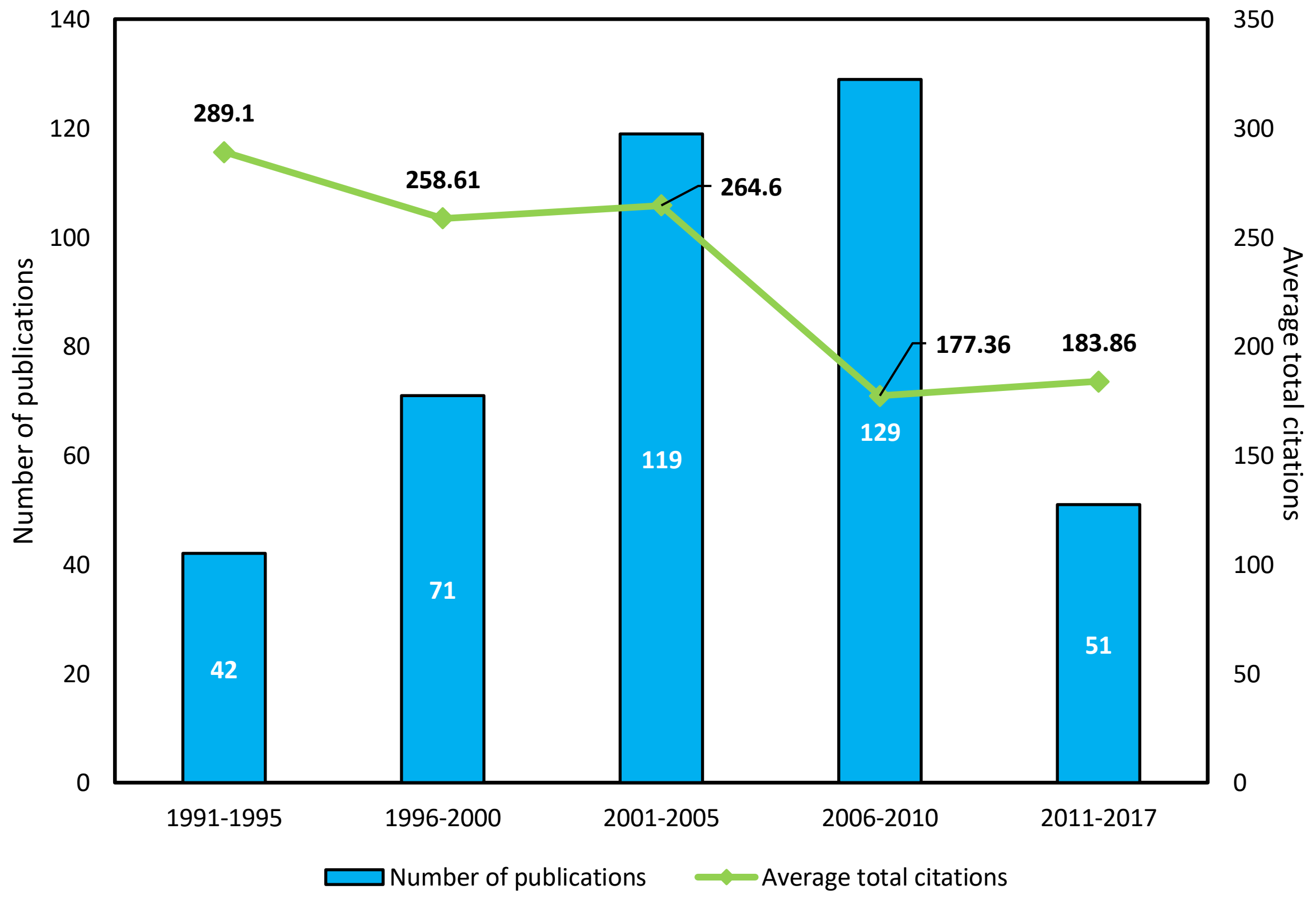

**Figure 2:** Number of publications and average total citations

After extracting all publications with total citations (TC) < 100, we obtain 412 highly cited publications – including 333 articles, 40 reviews, 33 proceeding papers, two editorial materials, two books, and two book chapters. The highly cited publications are written by 729 different authors, of which only 70 are authors of single-authored documents (less than 10% of total authors). Even though we retrieved data during 1970-2019, highly cited documents only exist during 1991-2017. We split the number of publications and their average citations into five timeframes for better visualization: 1991-1995, 1996-2000, 2001-2005, 2006-2010, and 2011-2017 (see Figure 2).

## 4.2. Ideological dominance

### 4.2.1. Author level

Y-index is an indicator that explores not only the relevance of an author but also his/her contribution characteristics (whether leadership or supervision). Here, we plot 17 most influential authors (barely more than 2% of total authors) in the field of entrepreneurial finance. Only authors acquiring a *j* score larger than five are qualified. As can be seen from Figure 3, most of the influential authors lie within the second area of the polar coordinate system. Only Cumming D and Zahra SA are located in the third and fourth areas, respectively.

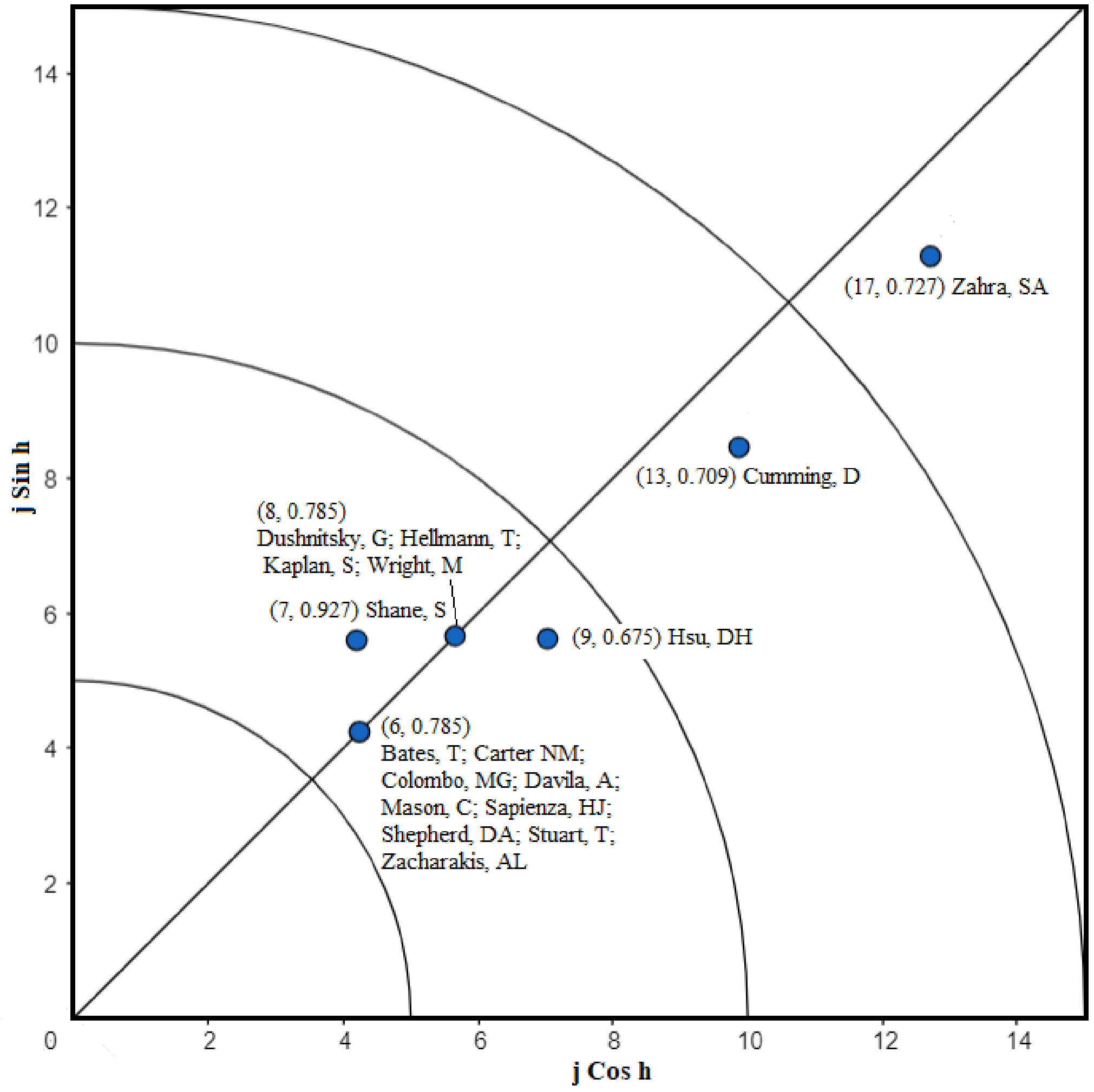

**Figure 3:** Distribution of top 17 authors who have $j \geq 5$

Zahra SA, with Y-index (17, 0.727), is the most influential author in the field of entrepreneurial finance, whereas Cumming D comes after with Y-index (13, 0.709). Both of them obtain *h* score of less than 0.7854, so they are more likely to take a leading position in a paper. Hsu DH also obtains a similar contribution tendency with Cumming D and Zahra SA. Out of 17 authors, 13 authors have the same number of *FP* and *RP*; thus, their *h* score is equal to 0.7854. Shane S, with *h* score higher than 0.7854, is the only author that holds a more significant number of *RP* than *FP.* To elaborate, Shane S is more likely to supervise the planning, execution of the study, and writing the article. Notably, all of 17 most influential authors are affiliated with institutions in Europe and North America.

### 4.2.2. Institution level

Among 371 recorded institutions, only 20 institutions with a *j* score higher than seven are selected (see Figure 4). The presence of institutions in the USA is dominant with 16 universities, while the other three institutions are from the UK (University of London, Imperial College London, and the University of Nottingham), and one is from Italy (Polytechnic University of Milan). All of 4 institutions outside the USA is located within the first area of the polar coordinate (from 0 to 20). While University of Nottingham and Imperial College London – Y-index (10, 0.588) and Y-index (11, 0.695) have reasonably high leadership tendency, University of London and Polytechnic University of Milan hold neutral position between supervision and leadership with Y-index (12, 0.785) and Y-index (8, 0.785), respectively.

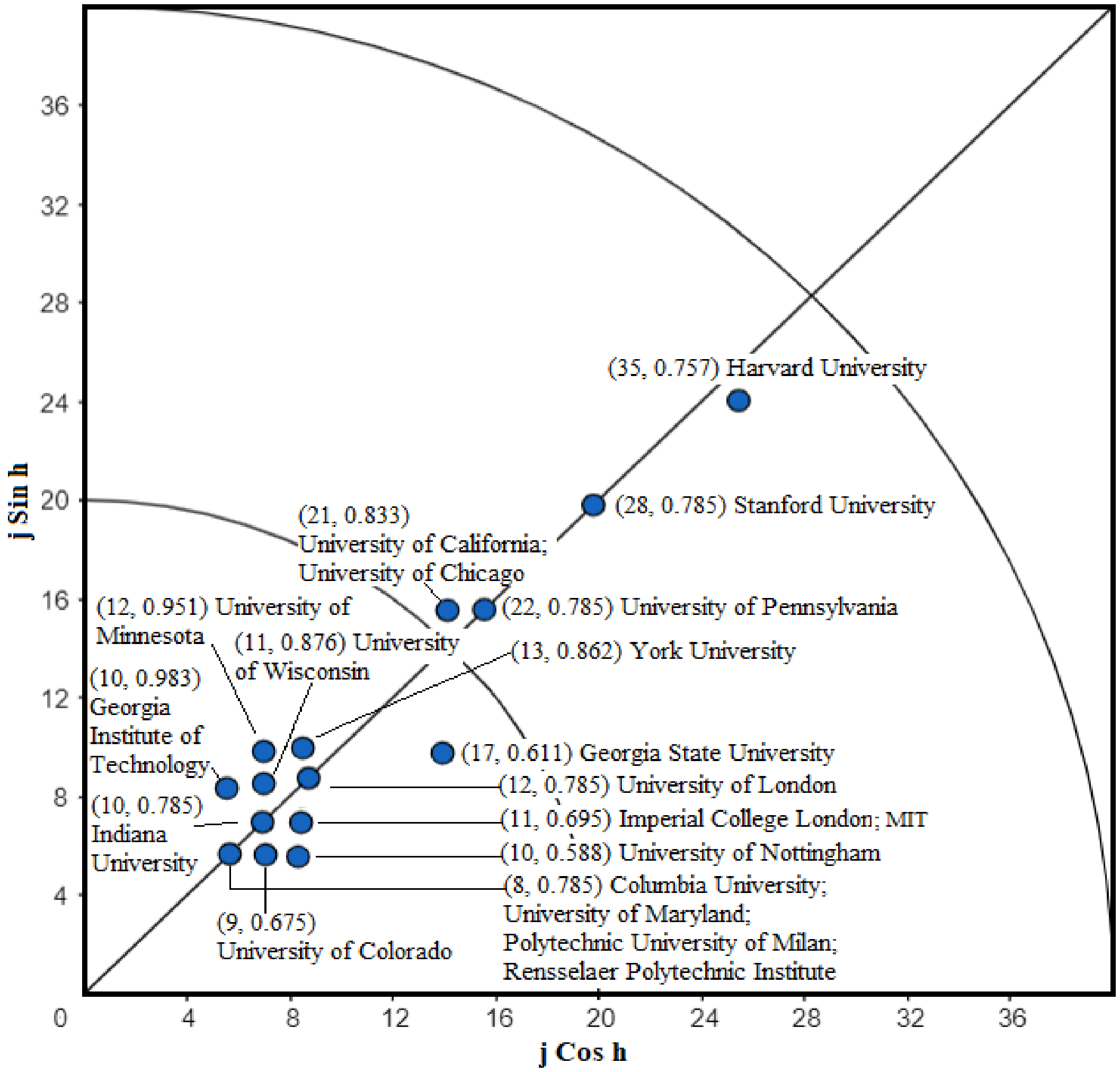

**Figure 4:** Distribution of top 20 institutions which have $j \geq 7$

The five most influential institutions in the field of entrepreneurial finance are all from the USA: Harvard University, Stanford University, University of Pennsylvania, University of California, and the University of Chicago. Harvard University is the only institution that has an *h* score of less than 0.7854, but the difference is negligible (*FP* = 18 and *RP* = 17). In contrast with Harvard University, the University of California and the University of Chicago – with Y-index (21, 0.833), are more prone to supervision. For other USA institutions in the first area of the polar coordinate, Georgia State University, MIT, and the University of Colorado are more likely to lead research (their *h* score is less than 0.7854). In contrast, York University, University of Minnesota, University of Wisconsin, and Georgia Institute of Technology are more likely to supervise or conceptualize research (their *h* score is higher than 0.7854).

With these results, we can see that the Western authors and institutions, especially those from the USA, substantially influence the ideology in entrepreneurial finance through the scientific output of most impactful publications.

#### 4.2.3. Country-level

At the national level, the Y-index presented in Figure 5 can be firm evidence for the Western ideological homogeneity as well as a sign of ideological hegemony of the USA in the field of entrepreneurial finance. Figure 4 depicts the Y-indexes of 15 most influential countries with the *j* score higher than seven. The USA, UK, and Canada are the three outliers with Y-index (511, 0.795), Y-index (85, 0.750), and Y-index (39, 0.811), respectively. Compared with the UK, the *j* score of the USA completely overweighs with a six-fold greater *j* score. Nevertheless, the USA is more prone to a supervision role in a paper (*FP* = 253 and *RP* = 258), while the UK is more prone to the leadership role in a paper (*FP* = 44 and *RP* = 41).

Looking at the blue box in Figure 4, we observe a notable trend. Western countries are more prone to supervision roles (such as France, Belgium, and the Netherlands) and neutral positions (such as Italy, Switzerland, Spain, Germany, and Australia) than a leadership role. In contrast, non-Western countries are more likely to hold a leadership role, especially China, with Y-index (10, 0.588).

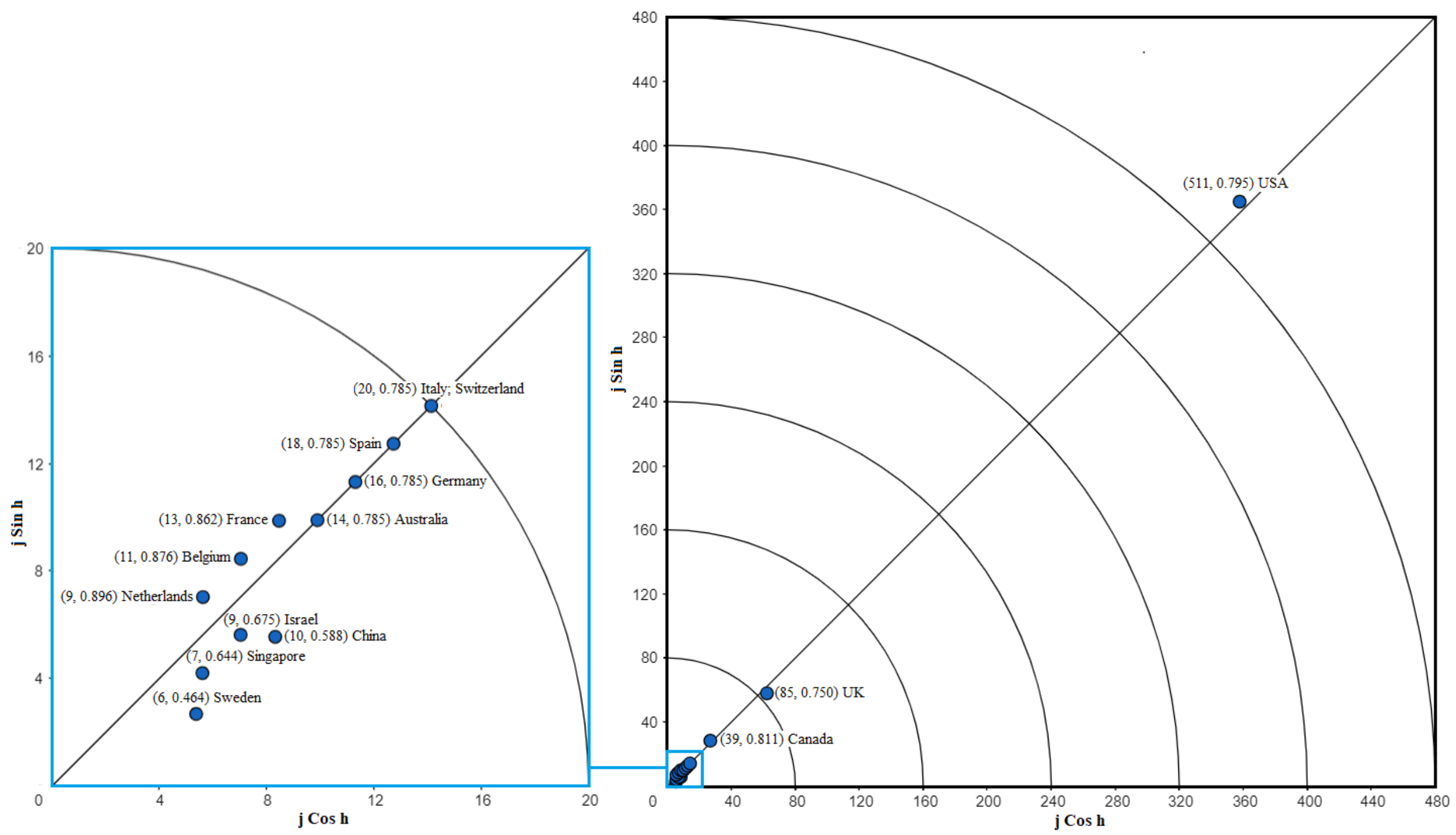

**Figure 5:** Distribution of top 15 countries which have $j \geq 7$

## 4.3. Ideological tolerance

### 4.3.1. Author level

At the individual level, the social structure of 729 authors is visualized employing the Louvain clustering algorithm and Kamada-Kawai layout (see Figure 6). We also set the minimum frequency between two authors as two, which means only the connections with the frequency being higher than one are included in the network.

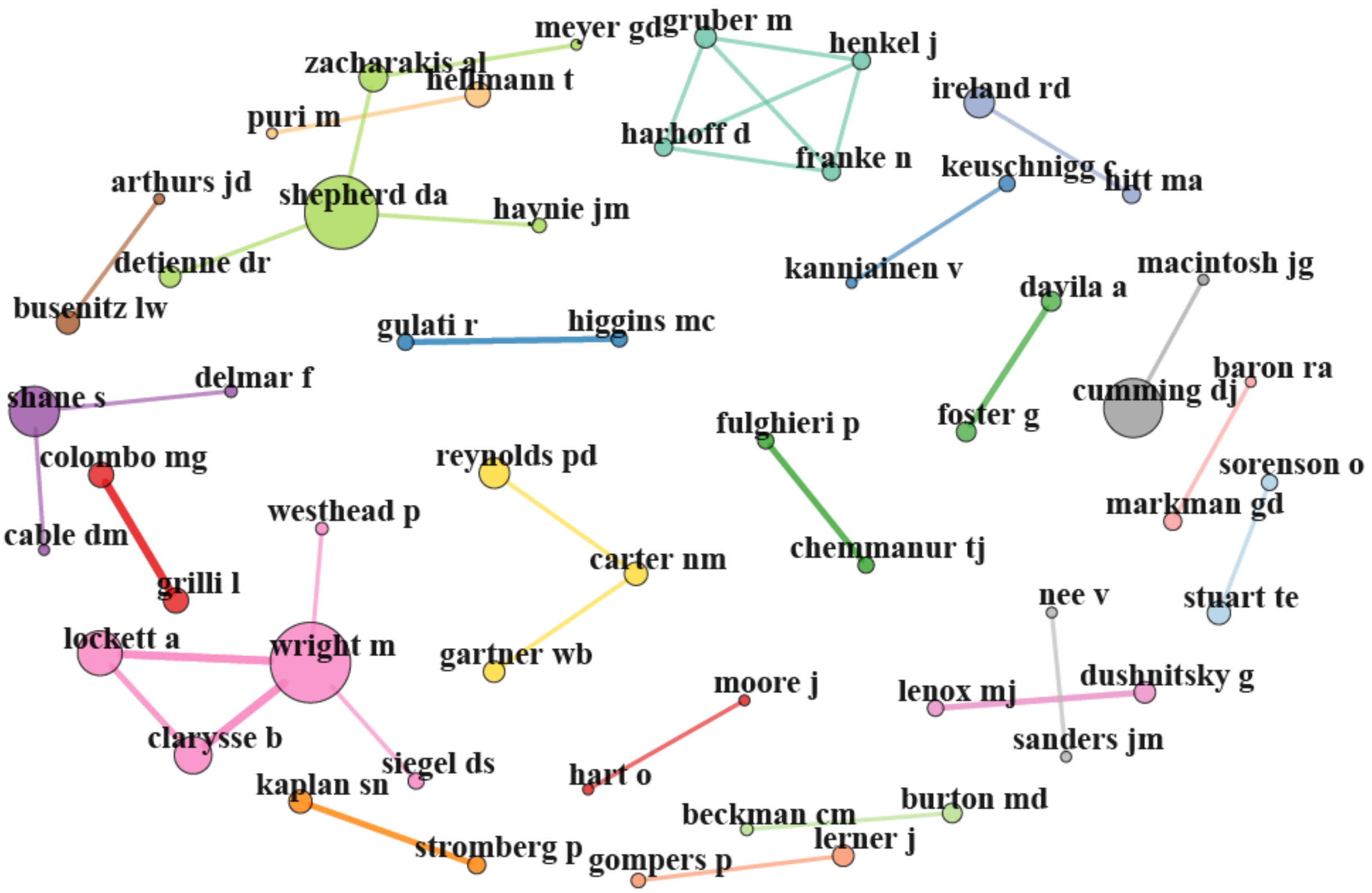

**Figure 6:** The social structure at author-level

In total, there are 22 research groups in the "nucleus" of entrepreneurial finance. The size of most research groups is relatively modest, with only two authors. The five authors with the highest number of collaboration links are Wright M (45 links), Shepherd DA (41 links), Zahra SA (40 links), Cumming DJ (33 links), and Shane S (28 links). In their research groups, other members are all from Western countries, mostly the USA. A similar collaboration pattern is also observed in other research groups in which members are from Western countries. It is plausible to say the tolerance level of heterogeneity at the author-level is weak, and entrepreneurial finance literature lacks knowledge exchange between top Western authors and non-Western authors.

#### 4.3.2. Institution-level

At the institution-level, we conduct the co-authorship analysis on 371 institutions using similar settings with the author level. Figure 7 display eight research networks within which member institutions publish at least two highly cited publications together. 27 Western institutions form these research networks; not-plotted institutions have either no collaboration with other institutions or collaboration with a frequency of less than two (see Figure 7). Among 27 institutions, top ten institutions with the highest number of collaboration links are in the USA; top five institutions are Stanford University (70 links), Harvard University (58 links), University of Minnesota (48 links), Babson College (47 links), and University of Chicago (47 links). Despite the high degree of collaboration, USA universities frequently collaborate with only institutions in Europe. The only international organization frequently collaborating with the US institution is the World Bank, but that collaborative connection is with a non-educational institution – the National Bureau of Economic Research. Again, ideological tolerance of non-Western values is also weak at the institution level.

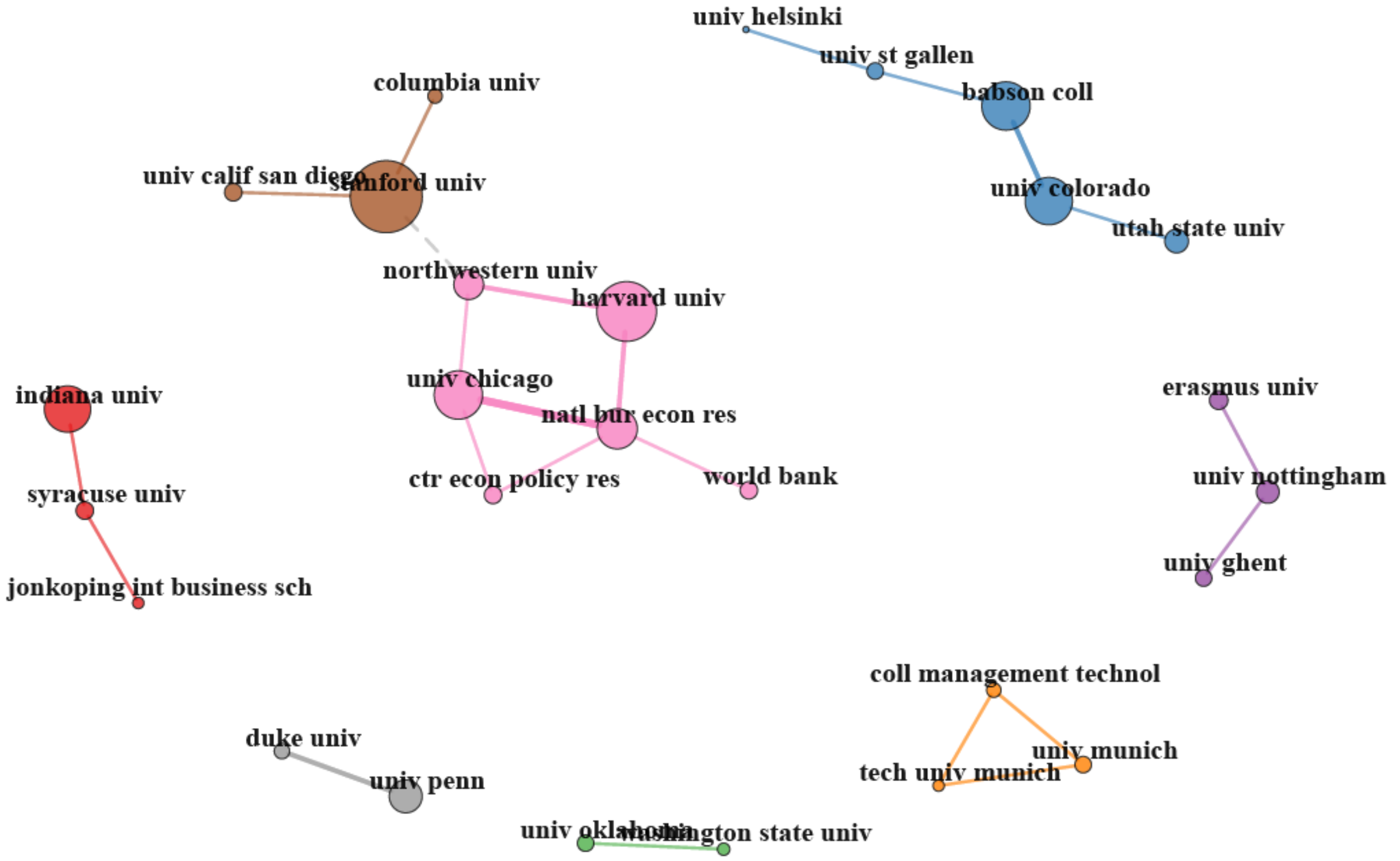

**Figure 7:** The social structure at institution-level

### 4.3.3. Country-level

Similar to the author- and institution-level, the social structure at the country-level is also plotted using the Louvain clustering algorithm and Kamada-Kawai layout. We also set the minimum frequency between two countries as two. Out of 32 analyzed countries, 19 countries are included in the collaboration system with three networks (see Figure 8). In this collaboration system, only China, Korea, and Brazil are non-Western countries, and they belong to the collaboration network led by the USA. European countries form the other two clusters. In terms of the total collaboration frequency, the US is dominant over other countries with 570 collaboration links, which is more than three times and a half of the second rank – the UK (158 links). At the country-level, there is a signal of knowledge exchange between Western and non-Western countries in the "nucleus" of entrepreneurial finance. However, the connection is very scant, given the global prevalence of entrepreneurship.

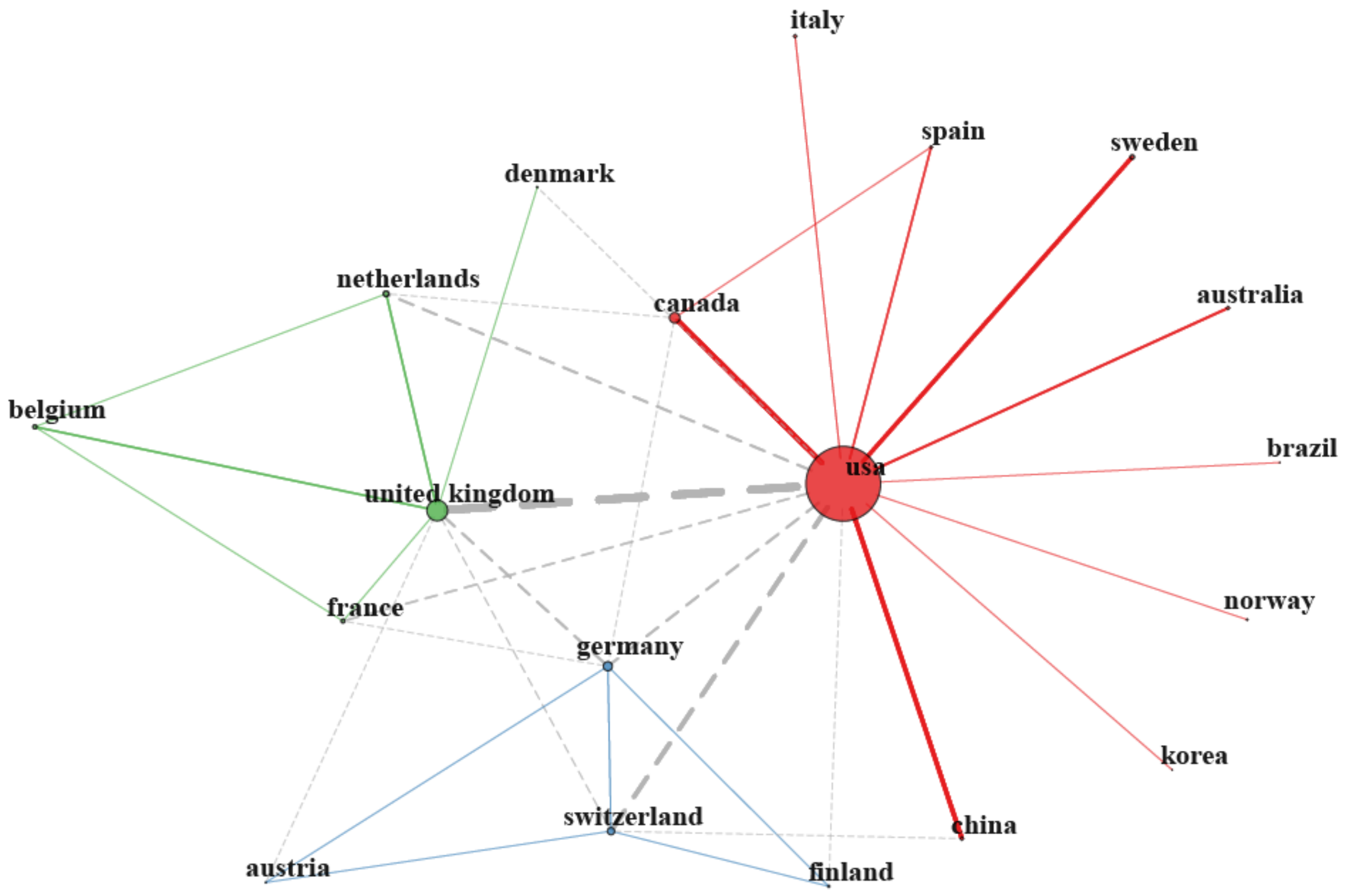

**Figure 8:** The social structure at country-level

## 5. Discussion

By calculating the Y-indexes across author-, institution-, and country-levels, we find the *dominance* of Western ideology, especially in the US, within the "nucleus" of entrepreneurial finance literature. Specifically, at author-level, all 17 leading authors in terms of Y-index ($j$ > 5) are from Western countries. At institution-level, 20 leading universities are Western-based, and 80% of them are located in the US. Several non-Western countries appear in the graph at the country-level, such as China with Y-index (10, 0.588), Israel with Y-index (9, 0.675), and Singapore with Y-index (7, 0.644). However, their influence is negligible compared to the US with Y-index (511, 0.795), the UK with Y-index (85, .750), and Canada with Y-index (39, 0.811).

Besides the *dominance* of Western countries, their weak *tolerance* toward heterogeneous ideologies is also observed. In the social structure within the "nucleus," Western authors, institutions, and countries are prone to collaboration with other Western counterparts. No frequent collaboration connection (collaborating more than one publication) between Western and non-Western authors/institutions is detected, despite the frequent connection between the National Bureau of Economic Research and the World Bank. At the national level, the USA seems to increase its tolerance towards non-Western countries like Brazil, China, and Korea. Still, the overall tolerance of non-Western ideologies is low.

Given the Western countries' *dominance* and weak *tolerance* of heterogeneity, we support the finding of Nguyen et al. (2020) that the entrepreneurial finance literature is Western ideologically homogenous. This situation needs to be changed because ideological homogeneity will consequently hinder the development of the discipline. In an ideologically homogenous community, views that challenge common knowledge or ideology are more likely to be rejected or ignored. Although scholars have the freedom to investigate any matter, ideological dominance influences the biases of reviewers and editors who play the roles as "trust evaluators" (see Figure 1), which subsequently prevents the dissemination of unconventional or unpopular knowledge. Thus, this dominance is more likely to make innovative or new ideas under-evaluated or suppressed (Atkeson & Taylor, 2019; Mahoney, 1977).

Second, in a research field with mostly "likeminded" core publications, the level of competitiveness varies between the minority who disseminate non-Western values and the majority who disseminate Western values. On the one hand, viewpoints of the minority tend to receive higher criticism, while the majority enjoy the favorable impact of commonly held perspectives and standards. Due to this seemingly "hostile" environment, the minority may be less willing to share their opinions. On the other hand, the minority may be encouraged to adjust their viewpoints toward the common core values that are set by the majority to avoid disagreement (Myers, 1975; Wojcieszak, 2010). Third, ideological homogeneity might undermine the creditability of science because scientific uniformity leaves blind spots and narrows the possibility to raise innovative intellectual inquiries (Duarte et al., 2015; Gray, 2019).

Based on the three possible adverse outcomes of ideological homogeneity, we recommend editors, reviewers, and authors to take more proactive attitudes in order to diversify knowledge in entrepreneurial finance. Editors, reviewers, and authors are "trust evaluators" that help filter unqualified values and build up a set of core values through peer-review and citation systems, respectively.

Is the predominance of venture capital and crowdfunding the reason behind Western ideological homogeneity? Or do non-Western authors obtain a higher rejection rate than Western counterparts because of sharing their local viewpoints? We cannot answer the second question, but we may know that a majority of editors and reviewers in top journals that have high credibility and accessibility are from Western countries. As Mahoney's (1975) 's experiment shows substantial prejudice of reviewers "against manuscripts which reported results contrary to their theoretical perspective," diversifying editorial and reviewer boards in top journals is crucial for providing a fairer filtering process for non-Western authors. Another measure top journals should implement to increase ideological heterogeneity is to initiate Special Issues that focus on financing sources in non-Western countries.

In addition, authors in non-Western countries also need to be proactive in pursuing research topics that are culture-based and advantage-based (Vuong, 2019). For example, entrepreneurs in Asia are prone to finance from family members, while governmental subsidies are crucial for entrepreneurs in Communist-led countries. Given the tremendous costs and risks to pursue non-common research topics in emerging countries (Vuong, 2018), researchers in developed countries, especially top authors, should be more open to collaborating with non-Western authors for inducing knowledge exchange and ideological diversity.

Our study is not without limitations. First, employing only data from WoS might lead to production and ideological biases, because social scientists of some countries (e.g., Japan) are prone to publish in the national database due to language barriers. Also, most of the journals in WoS are published in English. Second, the evidence provided by this study only indicates the ideologically homogeneous core values of entrepreneurial finance but does not consider publications in the "buffer zone" (non-highly cited publications), so emerging research trends with greater diversity cannot be detected. Third, the mindsponge mechanism of scientific discipline can be used to examine other facets of ideologies (e.g., political ideologies, etc.) as well as the chronological evolution of ideologies. Nevertheless, the current study only applies the mechanism on a fixed timeframe and ideological differences following geographical locations. Future studies are recommended to apply the mindsponge mechanism in various disciplines to study the chronological evolution and evaluate other types of ideological diversity.

## 6. Conclusion

Entrepreneurial finance is a rapidly growing field, but the matter of ideological homogeneity is raised for the sake of sustainable development of the field. This research is the first attempt to evaluate the ideological homogeneity in a scientific discipline by employing bibliometric techniques. Based on the Y-index and co-authorship analysis, we find Western ideological dominance and weak tolerance of heterogeneity among highly cited papers across three levels (author, institution, and country), which are strong evidence for the Western ideological homogeneity in entrepreneurial finance. Given various shortfalls of being ideologically homogenous, we recommend editors, reviewers, and authors to take proactive actions for diversifying research topics and enhancing knowledge exchange. Furthermore, the mindsponge mechanism can also be used to judge the chronological evolution and the diversity of core values/ideologies of scientific disciplines other than entrepreneurial finance.